\documentclass[a4paper,10pt]{article}                  
\usepackage{amsmath}
\usepackage{cite}
\usepackage{overcite}
\usepackage{graphicx}

\begin{document}
\twocolumn
\title{Polarization analysis of propagating surface\\ plasmons in a subwavelength hole array}
\author{\normalsize E. Altewischer, M.P. van Exter, and J.P. Woerdman \\ 
\it \normalsize Leiden University, Huygens Laboratory, P.O. Box 9504,\\ 
\it \normalsize 2300 RA Leiden, The Netherlands \\ 
\\
\small We present angle- and polarization-resolved measurements of the optical transmission \\
\small of a subwavelength hole array. These results give a (far-field) visualization of the \\
\small corresponding (near-field) propagation of the excited surface plasmons and allow for \\
\small a simple analysis of their polarization properties.}


\date{\today}
\maketitle 


\sloppy
\noindent
Metal films perforated with a periodic array of subwavelength holes have recently been shown to exhibit a much larger optical transmission than expected \cite{Ebbesen}. The reason for this is the presence of surface plasmons (SPs) on the metal interfaces, which couple to freely propagating light via the 2D-grating of the hole array. The incident light gets through because it first excites SPs on the front side of the metal layer, which then couple through the holes to SPs on the back side, and which finally reradiate in the form of light. Since its original discovery\cite{Ebbesen}, this phenomenon has been studied in many theoretical and experimental papers\cite{Martin,Ghaemi,Grupp,Salomon,Altewischer}. An aspect that has been underexposed up to now is the role of polarization in the transmission process. SPs are combined optical-electronic excitations that comprise a (longitudinal) electric field aligned with their propagation direction, which also determines their coupling to the (input and output) optical field. In this letter we address experimentally the polarization properties of the transmission, both by varying the polarization of the light incident on the hole array and by analyzing the polarization of the output light.

Our experiment differs from near-field microscopy studies on SPs\cite{Dawson,Sonnichsen,Hecht,Baida} performed so far. These experiments can map the optical field on a subwavelength scale, but have only been performed for smooth films\cite{Dawson,Hecht,Baida} or films with isolated holes\cite{Sonnichsen}. We have specifically studied the propagation and polarization of SPs on a metal layer with an {\it array} of holes, in which multiple interference of the SPs scattered from the holes is important, as this gives rise to SP-bandstructure. Moreover, we use the much simpler (polarization- and angle-resolved) analysis of the optical {\it far}-field from which this information can be easily extracted.

The sample that we have studied is a 1~mm~$\times$~1~mm hole array in a 200~nm thick gold layer, which is attached to a 0.5~mm glass substrate by a 2~nm titanium bonding layer. The metal layer is perforated with a square grid of 200~nm diameter holes spaced with a 700~nm lattice constant. A typical transmission spectrum for normal incidence on the hole array is shown in Fig.~1. The peak transmission of the most prominent resonance, 6\% at 810~nm, is much larger than the value of 0.56\% expected from ``classical" theory \cite{Bethe}. (This same resonance was recently used by us in an entanglement experiment\cite{Altewischer}.)
\begin{figure}
\centerline{\includegraphics[width=7.4cm]{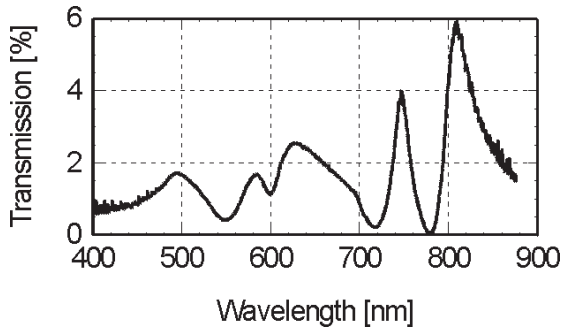}}
\caption{Wavelength-dependent transmission of the hole array used in the experiment, for unpolarized, normally incident (white) light.}
\end{figure}

The in-plane momentum of the SPs is supplied by the projected component of the incident optical wave momentum and a lattice component that is taken from the hole array\cite{Ghaemi}. Usually, when the surrounding media are not the same (air and glass, in our case), the observed transmission resonances are assumed to be dominated by SPs excited on either one or the other surface. On the basis of this simple model, we associate the 810~nm resonance with the four degenerate ($\pm1,\pm1$) modes on the metal-glass interface (for normal incidence).
Here we use the notation $(N_x, N_y)$ to label the SP resonance in terms of the lattice momentum needed for excitation, where the excited SPs propagate in the corresponding direction along the metal surface.  

The transmission spectrum depends on the angle of incidence, via the momentum of the excited SPs. A measurement of this angle dependence was used by Ghaemi et al. to determine SP dispersion curves \cite{Ghaemi} (i.e. their bandstructure); they measured the transmission spectrum over a large wavelength range for a set of fixed angles of incidence (for two orthogonal polarizations). (Elsewhere, we have reported such a measurement for our array\cite{Altewischer}). In contrast, in this Letter we use a fixed wavelength and monitor the transmission at this single wavelength as a function of the two far-field angles $(\theta_x,\theta_y)$ by using a CCD. This setup allows for a simple analysis of the polarization aspects of the transmission. For normal incidence (plane wave) these are trivial: the polarizations of input and output are the same, at least for a square array, as ours. This is because the incident light excites a frequency-degenerate set of SPs, each having a well defined polarization, which (vector) sum up to 'neutral' due to the square symmetry of the array. For instance, the set ($\pm N,0)$, which is x-polarized, is frequency degenerate with the set (0, $\pm N)$, which is y-polarized. For non-normal incidence this frequency degeneracy disappears and the polarization properties become anisotropic. As an aside, note that this would happen also upon going from a square to a rectangular array, but then even for normal incidence. 

To be able to observe SP propagation, the SPs have to be excited locally with an incident beam of light of sufficiently small dimensions, i.e. of the order of or smaller than the SP propagation length. Therefore, we have measured the far field of the transmitted light of the hole array by putting it in the focus of a telescope, consisting of a confocal configuration of two lenses with 15~mm focal length each. The entrance pupil was set by a 6~mm-diaphragm and was illuminated by a linearly polarized laser beam generated by an 810~nm wavelength Ti:Sapph-laser. We estimate the spot diameter at the focal plane as $\approx 2-3~\mu$m. The orientation of the input polarization was set with a halfwave-plate. A relay lens (10~cm focal length) was used to make a one-to-one image of the output telescope-lens on a CCD (Apogee), which therefore recorded the far-field of the array. In all images interference rings of equal inclination were visible. They result from the glass substrate acting as a Fabry-Perot and producing interference for the plane-wave components at different angles present in the focus. 

\begin{figure}
\centerline{\includegraphics[width=8cm]{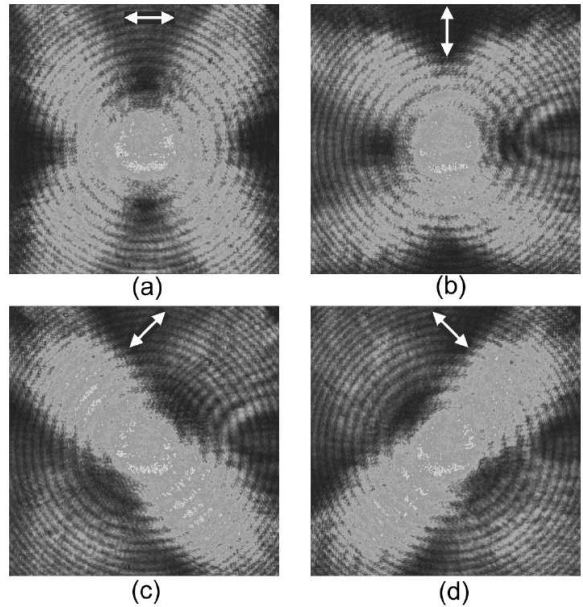}}
\caption{Far-field transmission of the hole array, at a wavelength of 810~nm ($17.6^\circ \times 17.6^\circ$ solid angle) as measured with a CCD. The input polarization (white arrows) was set at {\bf a} $0^\circ$, {\bf b} $90^\circ$, {\bf c} $45^\circ$, {\bf d} $-45^\circ$. The color scales used are different for each picture.}
\end{figure}
Fig.~2 shows the measured far-field of the ($\pm1,\pm1$) mode at a wavelength of 810~nm. Each of the four pictures was obtained for a different linear input polarization, (a) $0^\circ$, (b) $90^\circ$, (c) $45^\circ$ and (d) $-45^\circ$, indicated by arrows in the pictures. The throughput intensity is measured without any further polarization analysis. These pictures visualize the nature of the dominant SP resonances that are addressed at 810~nm. The prominent diagonal ''lobes" are consistent with the ($\pm1,\pm1$) labelling. For incident light polarized along $0^\circ$ (or $90^\circ$) the lobes on {\it both} the diagonals are excited. For $+45^\circ$ input polarization angle only the lobes along {\it one} of the diagonals are excited (c), for $-45^\circ$ the complementary ones (d). This shows that the incident optical electric field is decomposed into components along the four resonant modes, thereby exciting either one or both diagonals depending on the incident polarization angle. This is displayed schematically in Fig.~3 for incident light of 90$^\circ$ polarization angle, where the incident field (dashed arrow) is decomposed into two equal components along the modes lying on the diagonals (solid arrows).  

\begin{figure}
\centerline{\includegraphics[width=3.5cm]{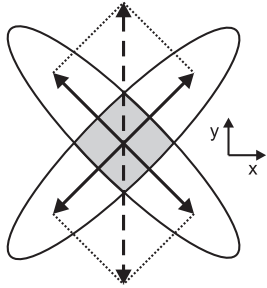}}
\caption{Schematic representation of the {\it near}-field at the backside of the hole array for 810~nm light. In the far-field, narrow and wide dimensions are interchanged due to Fourier relations, but the polarization directions remain unchanged. The incident optical electric field, polarized at 90$^\circ$ (dashed arrow), is decomposed into four resonant SPs (solid contour lines). These SPs propagate along, and are polarized at, $\pm 45^\circ$ (solid arrows). The shaded overlap region is polarization isotropic, thus polarized along the incident field.}
\end{figure}

Based on Fourier relations, a compact far-field in a certain direction corresponds to a wide near-field in that direction, as could occur due to SP propagation. This is evident in Fig.~2c, where an input polarization of $45^\circ$ gives a compact field in the same direction, whereas the orthogonal direction is wide. A typical value of this compact width is 4.1$^\circ$, which Fourier relates, assuming a Lorentzian far-field profile, to an estimated propagation length in the near-field of 4.3$\mathrm{\mu m}$ (FWHM). This value is much smaller than the literature value of 40 $\mu$m (at 800~nm) for a smooth gold surface \cite{Raether}, probably due to radiative decay induced by the hole patterning. 

In a second experiment, the polarization of the 810~nm mode was analyzed by putting a polarizer behind the hole array. Fig.~4a and b show the measured far-fields for an input polarization of $90^\circ$ (white broken arrows), thereby exciting both +45$^\circ$ and -45$^\circ$ modes equally (see also Fig.~3). In the first picture (a) the analyzing polarizer was set at $+45^\circ$ (red arrows). For this polarization, only the lobes corresponding to the (1,1) and (-1,-1) modes are visible. This shows that these modes are polarized at $+45^\circ$ and, at the same time, that the orthogonal (invisible) (-1,1) and (1,-1) modes are polarized at $-45^\circ$ (for which the polarizer is at blocking angle). In both cases, the polarization is in the same direction as the SP propagation direction, as expected. 
The second picture (Fig.~4b), made for analyzing polarizer at $0^\circ$ (blocking angle to input polarization), shows no intensity in the central overlap region (corresponding to the telescope axis), whereas the outward lobes (corresponding to off-axis) do have intensity. This is consistent with the central spot being polarization isotropic, corresponding to the shaded region in Fig.~3, where both orthogonal polarization-components are present and overlapping due to the finite width of the lobes (indicated by the solid contour lines). On the other hand, the outward lobes (outside the shaded region in Fig.~3) each have a unique polarization (indicated by arrows), because they are purely single-mode (i.e. non-overlapping). They behave quite similar to polarizers, as they transmit only the polarization component of the incident light aligned with their axis (at $\pm 45^\circ$). In the situation of Fig.~4b, where the hole array is effectively placed between two orthogonal polarizers (at $0^\circ$ and $90^\circ$), this means that the ''lobe-polarizers" (at $\pm45^\circ$) are expected to transmit 1/4 of the incident intensity, due to simple projection arguments.
\begin{figure}
\centerline{\includegraphics{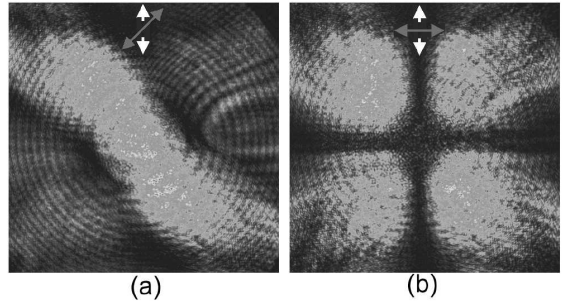}}
\caption{Polarization analyzed far-field transmission of the hole array (at 810~nm), with a polarizer in front of the CCD, oriented at {\bf a} $45^\circ$ and {\bf b} $0^\circ$ (red arrows). The input polarization is $90^\circ$ (white arrows).}
\end{figure}

A closer look at Fig.~2a and b reveals that they are not perfectly 4-fold symmetric, but have some 2-fold symmetry admixed (a and b are not fully identical, which should be the case for 4-fold symmetry). This slight symmetry breaking is not visible for the ''pure" lobes in Fig.~2c and d, which are intrinsically 2-fold symmetric. 
Fig.~4 shows again slight symmetry breaking, namely in the transmission values of the 4 outward lobes: off-axis, i.e. in the lobes, typical transmissions for $0^\circ$-input polarization are 35\% ($\pm 5\%$) for $0^\circ$-output polarizer, and 15\% ($\pm 5\%$) for $90^\circ$-output polarizer. Note that these transmissions are normalized to the maximum (center) value in Fig.~4a. These values differ significantly from the expected 25\%, based on the assumption that the transmission process is mediated by the ($\pm1,\pm1$) modes only. All the above observations demonstrate that, for certain details, it is too simple to consider only a single set of SPs living on one side of the array, i.e. only the ($\pm1,\pm1$) metal-glass modes. Theoretically, one expects some influence of the $(\pm1,0)$ and $(0,\pm1)$ modes on the air-metal side (at $\approx 750$~nm), as they are the nearest resonance. The admixture of these modes would be consistent with the observed 2-fold symmetry in the modal patterns in Fig.~2a and b and in the modal transmission values in Fig.~4b. 

On the basis of the SP assisted transmission-model, one expects that transmission resonances at different wavelengths have different far-field patterns and polarization properties. We have confirmed this by preliminary far-field measurements for the resonance at 750~nm. For this wavelength the patterns are more complicated than those for the ($\pm1,\pm1$) modes, which is related to the more complex dispersion curves at this wavelength; at the 750~nm resonance the modes of different ($N_x,N_y$) at different angles of incidence lie closer together than at the 810~nm resonance, and thereby have a stronger influence on the far-field patterns. However, the 750~nm resonance {\it does} show horizontal and vertical lobes, confirming the expected ($\pm1$,0) and (0,$\pm1$) nature of the SPs at this frequency. Elsewhere, we will present a more detailed study of our work, including the 750~nm resonance.

In conclusion, we have shown that polarization dependent measurements of the far-field are useful to analyze the optical properties of metal hole arrays; the polarization plays an important role, as it is related to SP propagation. In a way, SP modes act as polarizers that pass only the polarization component that is aligned with their propagation directions. Our measurements show that the gross features of the transmission can be adequately explained by a ''single-mode" SP model. The limitations of this simple model show up, however, indicating that more modes have to be taken into account for a realistic theoretical description of the transmission process. 

\section*{Acknowledgements} This work has been supported by the Stichting voor Fundamenteel Onderzoek der Materie (FOM); partial support is due to the European Union under the IST-ATESIT contract. We thank A. van Zuuk and E. van der Drift at the Delft Institute of Micro-Electronics and Sub-micron Technology (DIMES) in Delft, The Netherlands, for the production of the hole arrays.
\newline


\end{document}